# Harnessing Digital Twin Technology for Adaptive Traffic Signal Control: Improving Signalized Intersection Performance and User Satisfaction

Sagar Dasgupta, *Student member, IEEE*, Mizanur Rahman, Ph.D., *Member, IEEE, and* Steven Jones, Ph.D.

*Abstract*— **In this study, a digital twin (DT) technology based Adaptive Traffic Signal Control (ATSC) framework is presented for improving signalized intersection performance and user satisfaction. Specifically, real-time vehicle trajectory data, future traffic demand prediction and parallel simulation strategy are considered to develop two DT-based ATSC algorithms, namely DT1 (Digital Twin 1) and DT2 (Digital Twin 2). DT1 uses the delay experienced by each vehicle from all approaches connected to the subject intersection, while DT2 uses the delay of each vehicle that occurred in all the approaches connected to the subject intersection as well as immediate adjacent intersection. To demonstrate the effectiveness of these algorithms, the DT-based ATSC algorithms are evaluated with varying traffic demands at intersection, and individual user level. Evaluation results show that both DT1 and DT2 performs significantly better compared to the density-based baseline algorithm in terms of control delay reductions ranging from 1% to 52% for low traffic demands. DT1 outperforms baseline algorithm for moderate traffic demands, achieving reduction in control delay ranging from 3% to 19%, while the performance of DT2 declines with increasing demand. For high traffic demands, DT1 achieved control delay reduction ranging from 1% to 45% and DT2 achieved 8% to 36% compared to the baseline algorithm. Moreover, DT1 and DT2 effectively distribute the delay per vehicle among all the vehicles, which approach towards intersection, compared to the baseline ATSC algorithm. This helps to improve user satisfaction by reducing prolonged delays at a traffic signal, specifically, for moderate and high traffic demands.**

## I. INTRODUCTION

TRAFFIC congestion is a persistent challenge facing cities and transportation systems around the world that affects millions of people every day. Traffic bottlenecks often occur in city transportation networks, with a significant amount of delay experienced at intersections [1]. As cities continue to grow, it leads to longer commute times, increased fuel consumption, and higher levels of air pollution, which can negatively impact the environment and public health. According to the INRIX US National Signal Scorecards 2022 [2], traffic signals caused a delay of 10% in the travel time of an average trip, with each vehicle experiencing an average delay of around 18.3 seconds at signalized intersections. The study also reveals that the average total delay per signal reached 98.2 hours, resulting in a cumulative delay of approximately 23.7 million hours across the country. All these can have both short-term and long-term impacts on drivers, including increased levels of fatigue, decreased focus and concentration, and a greater risk of road accidents [3]–[7]. In addition, prolonged delays at a traffic signal can lead to feelings of anger, frustration, and annoyance, which can affect overall well-being and quality of life. There is no indication of mitigation of this problem whatsoever, instead, it is expected to only increase in the future because of increasing traffic demand and limited space availability of increasing number of lanes in urban areas.

A traffic signal control system (TSC) can mitigate delays at intersection by optimizing the signal timings based on traffic demand. There are three broad classifications of TSC: fixed time, actuated, and adaptive. Fixed-time or pre-times methods were the first traffic signal control techniques, in which the appropriate phase time for traffic signals is determined based on the historical data [8], [9] . Hence, Fixed-time methods may not be effective in managing the dynamic arrival rate of vehicles at intersections, as well as variations in traffic patterns on an hourly, daily, and weekly basis. Additionally, adverse weather conditions, such as rain, snow, and fog, along with unforeseen events like accidents or road maintenance, can further disrupt traffic flow, resulting in significant delays or longer vehicle queues [10], [11]. Actuated traffic control methods use sensors to detect the presence of vehicles, pedestrians, or bicycles at an intersection. Based on this detection, the controller can adjust the signal timing to optimize traffic flow and reduce congestion. However, the actuated controller does not optimize signal timing in real-time instead of use pre-optimized signal timing plans for different traffic demands. Adaptive Traffic Signal Control Systems (ATSC) system [12] is an advanced signal control technology that uses real-time traffic volume data to adjust the traffic signal timing dynamically, i.e., real-time. ATSC system relies on a network of sensors, such as loop detectors, cameras, and Global

Manuscript received xx xxxx xxxx; accepted xx xxxx xxxx. Date of publication xx xxxx xxxx; date of current version xx xxxx xxxx. This work was supported by the Alabama Transportation Institute under Grant xxxxxxxx.
Sagar Dasgupta is with the Civil, Construction, and Environmental Engineering Department, The University of Alabama, Tuscaloosa, AL 35401 USA (e-mail: sdasgupta@crimson.ua.edu)
Mizanur Rahman, Ph.D. is with the Civil, Construction, and Environmental Engineering Department, The University of Alabama, Tuscaloosa, AL 35401 USA (e-mail: mizan.rahman@ua.edu)

Positioning System (GPS), to collect real-time data related to traffic flow. This data is then analyzed by an algorithm that adjusts the signal timings based on the current traffic conditions. ATSC systems are shown to hold the potential to improve the performance of urban signalized intersection by reducing delays, travel times, and queues, thus generating additional capacity and improving the overall level of service (LOS) of the roadway systems [13], [14]. However, such improvements are limited by the sensor capabilities [15]. Recently, real-time signal optimization at signalized intersections is poised to undergo substantial changes with the availability of trajectory data utilizing connected vehicle (CV) technologies. However, these efforts are still in the nascent stages. As the industry slowly shifts towards intelligent or smart transportation systems, digital twin (DT) technology is expected to further revolutionize traffic management and operations by building upon the trajectory data and further analyzing it and offering relief to urban commuters by enhancing their travel experience.

A DT is a cyber-physical system (CPS) where two-way data exchange between digital and physical entities occurs in real time with the aim of improving decision-making. Steyn and Broekman [16] define DT as integrated multi-physics, multiscale, and probabilistic simulations of a complex product in manufacturing industries that mirrors the behavior and environmental responses of its corresponding physical twin. Rudskoy et al. [17] understand DT as a module that reproduces a detailed digital model of the road and allows for modeling and experiments to test solutions and simulate different situations. DT has been increasingly adopted in several fields, such as manufacturing and production engineering [18], [19], medicine [20], healthcare [21], systems engineering [22], and product design [23]. However, DT applications in transportation systems have thus far been limited and are still in the inception stages – and its application towards traffic signal control problems is currently non-existent. Transportation digital twin (TDT) is a digital representation of a transportation system or network, which can be created using data from various sources, such as cameras, GPS, and other monitoring systems. It creates virtual model that simulates the real-world transportation system and provides a platform to analyze and optimize its performance. The TDT is a dynamic model that can be updated in real-time with new data and can be used to evaluate the trade-off between various future scenarios and strategies [24]–[32].

By using TDT concept, it is possible to develop sophisticated ATSC systems that can proactively capture the complex interactions utilizing feedback loops between physical and digital world due to the availability of precise vehicle-level information. It is important to note that a requirement of DT representation of an object is that there must be real-time synchronization of the virtual and physical objects through a frequent update of the state information related to static (e.g., traffic signal controller) and moving (e.g., vehicle) objects in a transportation system. The models and simulations within DT systems consume real-time or near real-time data from physical sensors. The data from these sensors allow to update the simulations using real-time traffic demand as well as capture the evolution of traffic patterns and evaluate the trade-off between different future traffic demands proactively thereby. A DT system can incorporate a digital representation of the physical asset, but it must be augmented by the injection of real-time data, and parallel simulations needs to be performed using different future traffic demands to determine the tradeoff between different ATSC algorithms to meet the user satisfaction at a signalized intersection. It is our hypothesis DT-based ATSC could not only reduce the amount of time drivers spend waiting at traffic signals, but also redistribute the delays between travelers from different approaches to improve users' satisfaction. Moreover, it could help to reduce the likelihood of accidents caused by driver frustration or impatience. Overall, TDT-based ATSC is a proactive ATSC approach, which uses all possible future traffic demands at an intersection, utilizes those traffic demands to run multiple parallel simulations in real-time, and conducts a tradeoff between outcomes from all possible simulations in terms of delay reduction and distribution of delay among all users to find and implement the best TSC algorithm.

In this study, a TDT technology based ATSC system framework is presented for improving signalized intersection performance and user satisfaction. Specifically, real-time individual vehicle trajectory data, future traffic demands, and parallel simulation strategy are considered to develop two DT-based ATSC algorithms, namely DT1 (Digital Twin 1) and DT2 (Digital Twin 2). DT1 uses the delay of each vehicle from all approaches related to the subject intersection, while DT2 uses the delay of each vehicle that occurred in all the approaches related to the subject intersection as well as immediate adjacent intersection. To demonstrate the effectiveness of these algorithms, the DT-based ATSC algorithms evaluated with varying traffic demands (i.e., low, medium, and high) at two levels: the intersection level, encompassing intersection and approach performance, and the individual user level, delving into the realm of user experience. There are two primary contributions of this study in the TDT field as follows:

- Firstly, we introduce a comprehensive TDT-based ATSC system framework, which is the first of its kind. The proposed framework enhances conventional signal control proactively by adapting timings to accommodate changing traffic patterns during various periods. We introduce the concept of parallel simulation strategy for different traffic demand to assist in selecting the best algorithm for the intended application.
- Secondly, we present two TDT-based ATSC algorithms designed so that for various traffic demands, i.e., low, medium, and high traffic demand, respectively. These algorithms aim to reduce delays, improve user satisfaction, and enhance the overall efficiency at a signalized intersection.

The reminder structure of this paper is outlined in this paragraph. The paper consists of six sections. The second section provides a review of the existing literature related to ATSC and TDT, identifying gaps in the literature that the study aims to address. In Section 3, the TDT-based ATSC framework is presented in detail, including the different components and functions of the TDT environments and the DT1 and DT2 algorithms. The fourth section discusses the evaluation of the proposed framework and provides an in-depth description of the traffic simulation setup, including the baseline algorithm and evaluation metrics used. Section 5 presents the performance evaluation of the DT1, DT2, and baseline algorithms at the intersection, approach, and user levels. In Section 6, future research directions are proposed on how the TDT-based ATSC framework can be implemented in a city-wide network. Finally, Section 7 concludes the paper by summarizing the key findings and identifying potential research topics for future studies.

## II. Literature Review

Researchers have turned towards ATSC models [12] because of the variability of traffic parameters and the nonlinearity of this fluctuation. Several types of machine learning models have been developed for ATSC systems throughout the past decade [33]–[41]. The bulk of current articles selects average journey time, delay time, and intersection queue length as their optimization objective functions. The effectiveness of all of these models is limited due to various factors, such as insufficient and skewed data, lack of real-time data, complex and dynamic traffic conditions, and the computational and time-intensive nature of those techniques. The addition of real-time information gathering, along with advanced data analytics and the proliferation of machine learning algorithms is likely to benefit the future of transportation [42]. Predictive TDT models have the ability to anticipate traffic flow issues, such as congestion, bottlenecks, accidents and determine the required modifications to the TSC system in order to prevent or minimize the negative consequences [43], and the practical feasibility of such proactive models has been verified [44]. Besides, the potential to enable autonomous vehicles (AVs) as a crucial component of TDT has also been documented [45].

The TDT concept is still evolving. A summary of TDT-related research is presented in Table 1. Based on Table 1, it is evident that DT technology has become an important area of research in the field of transportation. The majority of the studies focus on the development of DTs for various transportation-related applications, including driving assistance, safety improvement, and mobility services. One of the applications of DT technology in transportation is for driving assistance systems. [46] developed a DT architecture for cloud-based cyber-physical systems, which was used to create a prototype for an advanced driver assistance system. [24] also developed a vehicle-to-cloud-based advanced driver assistance system using DT technology. Several studies have focused on using DT technology to improve safety in transportation. [25] and [47] developed DT-based visual guidance systems that integrate camera images with cloud-based knowledge to improve driving safety. [48] and [49] used DT technology to predict potential future actions of neighboring vehicles, enhancing connected vehicle safety. [49] developed a road system DT that explored the potential of innovative and intelligent technologies for all aspects of the road system, including the road, tire, and vehicle, to address the challenges of future mobility. Another area of focus in DT research is mobility services. [50] developed a simulation architecture for DT utilizing the Unity game engine to optimize connected and automated vehicle operation and safety applications. [30] designed a mobility DT framework based on an artificial intelligence-based data-driven cloud–edge–device framework for mobility services. In addition, some studies have explored the potential of DT technology for optimizing the charging scheduling and navigation algorithm for electric vehicles [51] and developing a standardized framework for vehicular DTs that enables efficient data collection, processing, and analytics phases [52]. Recently, [32] developed learning-based algorithms for personalized behavior modeling and online lane change prediction, aiming to aid connected and automated vehicles to drive like human-driven vehicles and improve their user acceptance and trust. The integration of DT technology with the ATSC system to create a real-time traffic decision-making is one of the areas where research is currently lacking. The integration of DT technology with the ATSC system to create TDT has the potential to provide dynamic adjustments to green time proactively based on real-time traffic data, which can effectively mitigate traffic congestion and enhance user satisfaction. Therefore, the focus of this study is to explore the potential benefits of TDT-based ATSC for managing traffic flow in urban roadway networks.

TABLE 1
SUMMARY OF EXISTING RESEARCH ON TDT

| Reference | TDT Focus | TDT Application | Approach |
|---|---|---|---|
| [46] | Vehicle | Driving assistance application | Development of a DT architecture for cloud-based CPS, along with a prototype for an advanced driver assistance system. |
| [53] | Vehicle, Driver | Frequent traffic congestions mitigation | Predicting driver intention. |
| [48] | Driver | CV safety | Predicting potential future actions of neighboring vehicles. |
| [51] | Vehicle, Electric vehicle, Infrastructure | DT simulation platform to optimize the charging scheduling and navigation algorithm for EV | Developing a simulation platform to model the entities of the smart grid such as mobile EVs and charging piles. |

| Reference | TDT Focus | TDT Application | Approach |
|---|---|---|---|
| [24] | Vehicle | Advanced Driver Assistance System | Developing a vehicle-to-cloud-based advanced driver assistance system. |
| [25] | Vehicle | Potential risk avoidance | Developing a DT-based visual guidance system by implementing a novel sensor fusion methodology, integrating camera image from the cloud. |
| [47] | Vehicle | Driving safety improvement | Developing a visual guidance system using a novel sensor fusion methodology that integrates camera images with cloud-based knowledge from the DT. |
| [54] | Vehicle, Infrastructure | Cooperative driving for safer, efficient, and economical driving | Developing an infrastructure-vehicle cooperative autonomous driving. |
| [55] | Roads' infrastructure | DT of road infrastructure creation | Developing a methodology to establish a DT of road infrastructure. |
| [56] | Road-Tire-Vehicle | Road safety improvement | Developing a road system DT, which explores the potential of innovative and intelligent technologies for all aspects of the road system, including the road, tire, and vehicle, to address the challenges of future mobility. |
| [50] | Vehicle | CAVs operation and safety application | Developing a simulation architecture for DT utilizing the unity game engine. |
| [57] | Vehicle | ITS implementation | Developing a DT architecture for the implementation of various ITS services. |
| [52] | Vehicle | AVs safety and security | Developing a standardized framework for vehicular DTs that enables efficient data collection, processing, and analytics phases. |
| [58] | Vehicle and Driver | Ramp merging assistance | Sharing advisory information with mainline and ramp vehicles. |
| [49] | Vehicle | AVs safety | AV's perception enhancement beyond the limits of its onboard sensors using ITS. |
| [30] | Human, Vehicle, Traffic | Artificial intelligence-based data-driven cloud–edge–device framework for mobility services | Developing a mobility DT framework. |
| [28] | Human, Vehicle, Traffic | Automated Driving System Testing, CPSs, Parallel Driving, Safety Critical Services, Traffic Management Centers, Digital Maps, Onboard Diagnostics, Logistics | Understanding DTs and the role they play in the evermore complex landscape of CAV modeling and simulation. |
| [59] | Vehicle | Safety and mobility improvement | Designing a cooperative driving system for non-signalized intersections, in which CVs collaborate with each other to traverse intersections without coming to a complete stop. |
| [32] | Vehicle | Train CAVs to drive like human-driven vehicles, and improve their user acceptance and trust | Developing learning-based algorithms for personalized. behavior modeling and online lane change prediction. |
| Our Paper | Vehicle, Traffic signal controller | Congestion mitigation, road user experience enhancement | Developing two DT based ATSC algorithms. |

Note: CV=Connected Vehicle, CPS=Cyber-Physical Systems, DT=Digital Twin, EV=Electric Vehicle, CAV=Connected and automated vehicle, ITS= Intelligent Transportation System, AV=Autonomous Vehicle, ATSC=Adaptive Traffic Signal Control

## III. TDT-based ATSC Framework

In this section we have presented a conceptual, yet comprehensive TDT based ATSC framework, two TDT based ATSC algorithms, and a framework for implementing the presented algorithms.

### *A. Conceptual Framework*

The framework presented in this study makes a direct contribution to clarifying how Digital Twin (DT) technology can be effectively implemented in the transportation domain. Merely creating a real-time digital replica of physical entities (PEs), such as cars and pedestrians and visualizing them is not enough to fully leverage the efficacy of DT. The true strength lies in the ability to analyze vast amounts of real-time data and make proactive, real-time decisions based on that data. However, the question remains: how can this be accomplished? Simulating scenarios and adjusting PE states at every timestamp is not feasible, as the ground truth states of the PEs and those of the simulation may not change in the same manner. This is a significant challenge in the research community, and the framework proposed in this paper provides a viable solution to address this issue.

Figure 1 showcases a fundamental and comprehensive DT-based ATSC framework, which is a first-of-its-kind. Prior to this work, Qi et al. and Tao et al. [60], [61] proposed a five-dimensional DT model and Jones et al. [62] presented an eight-dimensional DT model. However, the TDT-based ATSC framework introduced in this paper surpasses these previous models by having nine dimensions. The nine dimensions can be expressed as shown in Equation (1).

$$ATSC_{TDT} = (PE, DS, DD, MO, SI, TP, CA, AP, CG) \qquad (1)$$

where, PE are physical entities, DS are digital shadow or replica of PE, DD denotes data, MO stands for models, SI stands for simulations, TP stands for traffic demand prediction, CA stands for TSC algorithms, AP denotes different ATSC algorithms and CG stands for communication gateway. Both Tao's [60], [61] and our proposed model share common components, including PE, DS, AP, DD, and CG. However, our model incorporates unique elements, including MO, SI, CA, and AP, and specifically, CA and AP are exclusive to the TDT-based ATSC framework.

The presented framework comprises of three layers: (i) Physical World; (ii) Digital Twin; and (iii) Communication Gateway. The physical world serves as the bottom layer and encompasses all transportation related physical objects, such as traffic signals, vehicles, road network, roadside infrastructure, and pedestrians. The top layer represents the DT which consists of digital replica of the physical world, database, models, traffic control algorithms, traffic demand prediction models and the ATSC applications. The top and bottom is connected by the communication gateway facilitating real time data exchange between these two layers.

The presented framework is an end-to-end framework similar to mobility DT framework proposed by Ziran et al. [30]. The two ends are sensing and actuation, and both of them are major functions of the physical world. The physical world includes both static and dynamic elements. The static elements refer to those objects, such as the road network, traffic signals, and roadside infrastructure, which are characterized by their stationary location. On the other hand, the dynamic elements refer to those objects, such as vehicles, pedestrians, and other road users, which are characterized by their frequent changes in location. Vehicles are equipped with sensors, such as positioning (i.e., GPS), inertial sensors (speedometer, accelerometer, gyroscope), camera, RADAR, LiDAR and sensed data related to vehicles motion states, such as speed, acceleration, heading and the data regarding their surroundings can be streamed to the DT layer. Roadside infrastructures are equipped with computing devices, camera, RADAR, LiDAR and other sensors to detect and monitor traffic volume, speed direction as well as to detect pedestrians and people using non-motorized vehicles [63]–[66]. Along with transmitting sensor data, health of the sensors is also transmitted to keep a track of active and inactive sensors. To ensure that the sensed data accurately reflects the state of the PEs, it is necessary that the data are of high quality and transmitted with minimal packet loss [67]. Data can be transmitted through either wired or wireless technology. While a wired connection may be suitable for stationary elements, wireless connectivity is necessary for dynamic elements. Ethernet is the prevailing technology for wired communication, while wireless communication technologies such as Wi-Fi, Cellular, Radio, Zigbee, and Long-Range Radio (LoRa) are widely adopted [68]. The feedback from ATSC applications module of the DT layer is transmitted back to the physical layer, and the PE actuate in response to the feedback.

DT layer consists of DS, digital sibling and ATSC applications. In the context of TDT, DS is the digital replica or DTs of the PEs i.e., vehicle, pedestrian, traffic signal, road network and roadside infrastructure. It is created and updated in real-time aggregating, synchronizing, fusing the data transmitted from the physical world [69]. The DT leverages data from the surrounding environment of the vehicles and roadside infrastructure to extract accurate traffic conditions, behavior, and movement of physical entities. In order to reduce storage requirements, the DS module temporarily stores raw data collected from the physical world. The communication between DS and the physical world is unidirectional. DS receives sensor data from PEs and stores processed data in the digital sibling database, but it cannot transmit information back to the physical world. Digital sibling has five basic components: (i) Database, (ii) Model, (iii) Traffic signal control algorithms, (iv) Traffic demand prediction models, and (v) Simulation. Real time data from digital shadow are stored in the database along with simulation results. These databases can be stored in multiple data lakes and data warehouses. Real-time data that includes the location and state of vehicles, pedestrians, and the state of the traffic signals are collected from the DS, stored, and processed in real time. Road network data and roadside infrastructure location and activity status data are stored as historical data. Whenever there is a change in the road network or roadside infrastructure, the database is updated accordingly. Historical data refers to information that has been collected and stored over a period of time, typically for reference or analysis purposes. It includes data such as traffic volume, traffic signal phase, waiting time, or vehicle trajectory data from GPS devices. Historical data can be used to identify trends and patterns in transportation behavior and feed to the simulation model. Models for simulating and predicting traffic flows comprise the model's module. Macroscopic, microscopic, and mesoscopic models are used for the traffic flow simulation. In the macroscopic model, traffic behavior is simulated at a high level which is useful for predicting overall traffic patterns and congestion levels but may not provide detailed information on individual vehicles. Whereas, in the microscopic traffic flow model factors such as vehicle speed, acceleration, and deceleration are considered to create a detailed representation of individual vehicle behavior and are used to simulate complex traffic scenarios, such as merging and lane changing. The mesoscopic traffic model falls between microscopic and mesoscopic models where road networks are divided into larger sections, such as roadway segments or clusters of intersections, and the flow of traffic between these sections is simulated. Traffic flow models consist of several components, including lane-changing models, driver behavior models, traffic control models, car following models, etc. In the ATSC application module, the traffic simulation output is utilized by the ATSC applications for decision-making processes, including algorithm selection and expected outcomes of different ATSC algorithms. The simulations are conducted based on predicted traffic demand since the implementation of ATSC algorithms is in the future. The traffic demand prediction module uses real-time traffic demand data from the database and prediction models from the model section. Traffic signal control algorithms are utilized for parallel simulations to identify the most suitable control algorithm for traffic flow optimization. The simulations utilize information, models, and algorithms, and the resulting output is fed back to the ATSC applications.

demands, the best performing ATSC algorithm from the simulations is utilized.

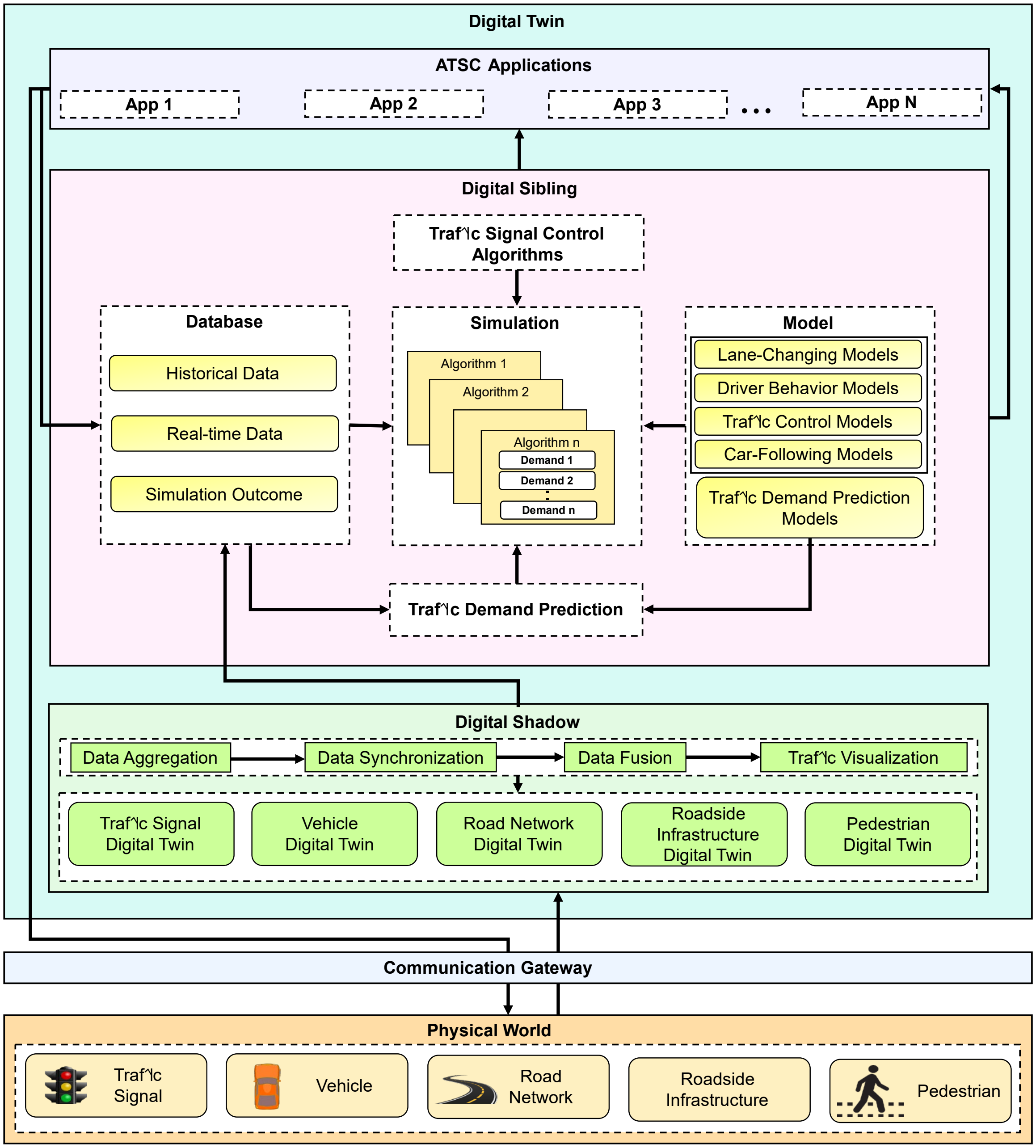

**Fig. 1.** Illustration of the TDT based ATSC framework.

The conceptual innovation in this framework lies in addressing the central question posed at the beginning of this sub-section. While the core components of the presented TDT framework are adapted from conventional DT frameworks to support ATSC applications, the addition of the simulation module is a critical feature. In this module, multiple simulations are initiated in parallel for a given timestamp, with initial conditions based on predicted traffic demands, and multiple ATSC algorithms are applied to determine the best-performing algorithm for each demand. When the traffic demand in a future timestamp matches one of the predicted

### *B. TDT-based ATSC Algorithms and Implementation Framework*

This section presents two delay based ATSC algorithms and the corresponding TDT framework. It also serves as a proof of concept for the TDT based ATSC framework presented in the previous subsection. Let's assume each subject intersection has eight connected approaches, four of which are through approaches and the remaining are dedicated left approaches as shown below (see Figure 2). These approaches are labeled as follows: east bound through (EBT), west bound through (WBT), north bound through (NBT), south bound through (SBT), east bound left (EBL), west bound left (WBL), north bound left (NBL), and south bound left (SBL).

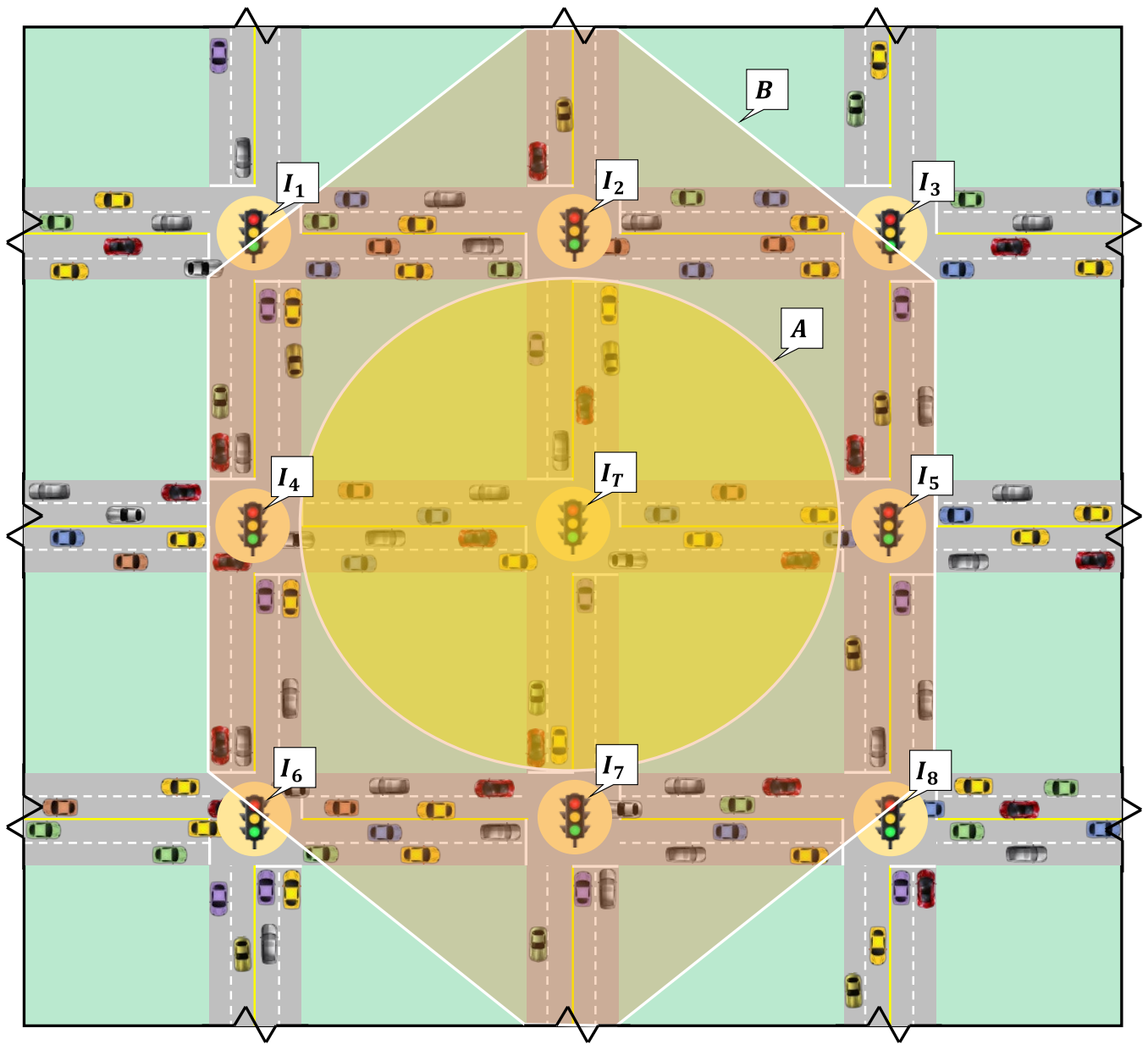

**Fig. 2.** Subject and connected intersections.

The pseudocodes of the traffic signal control algorithms DT1, and DT2 are presented below. In all the algorithms traffic phase change decision is evaluated every five seconds. A phase change is warranted when the proposed or set phase is different than the current phase. When warranted, the phase change follows through a yellow phase of two seconds and an all-red clearance of one second. Figure 3 illustrates the signal phases and their corresponding states. For instance, if the current phase is 0 and the proposed state is 6, the phase will first change to 1 (yellow phase) for two seconds, then to 8 (red phase) for one second, and finally, the proposed phase 6 is executed. A phase change is not warranted if the approach currently serves green and continues to be maximum after 5 seconds. In DT1 and DT2 algorithms, the green phase is assigned to the approach with the maximum average approach delay. For example, the algorithm will give the green phase to the EBT approach if its average approach delay is the highest among all approaches. If the signal is out of order, it will show flashing yellow. However, the definition of approach delay is different for DT1 and DT2. Generally, the approach delay is the delay experienced by the vehicles upstream of the intersection and it includes the delay due to acceleration and deceleration as well as the time the vehicle is fully stopped. Whereas in this paper the approach delay is calculate based on the stopped delay. Stopped delay is the time vehicle is stopped or moving very slowly. It is commonly utilized as a metric to indicate the degree of traffic congestion or the effectiveness of traffic management systems. The definition of "moving very slowly" is not well established. (Ahn et al., 2023) defined stopped delay the duration in which a vehicle's speed remains below that of a pedestrian, which is set at 1.2 ms-1. In this study stopped delay is defined as the duration in which a vehicle's speed remains below 0.1 ms-1. The threshold of 0.1 ms-1 is chosen to match it with the traffic flow simulator used in this study. The details of the stopped delay calculation are presented in the Evaluation setup section.

The average approach delay for DT1 is calculated using the accumulated stopped delays of vehicles on the approach that is connected to the subject intersection. On the other hand, for DT2, the accumulated stopped delays of vehicles on the approach connected to the subject intersection as well as the immediate previous intersection are used. Hence, DT2 utilizes more detailed information about individual vehicles. The green phase is warranted based on the average of the accumulated stopped delays of the connected approaches. For example, in Figure 2, the traffic signal controller of the subject intersection $I_T$ decides the green phase based on accumulated stopped delays experienced by the vehicles within the yellow circle in DT1. Whereas in DT2, accumulated stopped delays of the vehicles experienced within the apricot hexagon is used. For east bound through approach, the subject vehicles are all the vehicles between intersection $I_T$ and $I_4$. The average approach delay for east bound through approach, $D_{app\ EBT}$ is calculated using the following equation.

$$D_{app\ EBT} = \frac{\sum D_{veh\ EBT\ i}}{n} \quad (2)$$

where, $D_{veh}$ is vehicle's accumulated stopped delay and $n$ is the number of vehicles.
If $I_T$ implement DT1 algorithm, then the average approach delay will be the average of the all the accumulated stopped delay of all the subject vehicles. The delay of subject vehicle $i$ for DT1, $D_{veh\ DT1\ EBT\ i}$ is calculated using the following equation.

$$D_{veh\ DT1\ EBT\ i} = D_{veh}(t_{current}) - D_{veh}(t_{in}) \quad (3)$$

where, $D_{veh}(t_{current})$ is the accumulated stopped delay recorded in the vehicle at current time stamp and $D_{veh}(t_{in})$ is the accumulated stopped delay recorded in the vehicle at the time of entering the subject approach.

The approach delay of subject vehicle $i$ for DT2, $D_{veh\ DT2\ EBT\ i}$ is calculated using the following equation.

$$D_{veh\ DT2\ EBT\ i} = D_{veh}(t_{current}) - D_{veh}(t_{in}) + D_{veh}(I_{T-1}) \quad (4)$$

where, $D_{veh}(I_{T-1})$ is the accumulated stopped delay carried over by the vehicle from the previous approach.

Figure 4 presents the ATSC framework to implement the ATSC algorithms presented in this section. The physical world comprises of the road network, vehicles, and traffic signals. The physical world comprises the road network, vehicles, and traffic signals. The digital twin layer generates digital replicas of the physical entities by aggregating, synchronizing, and fusing data from the physical world. The components of the digital twin are the same as those presented in previous sub-section. A real-time traffic signal control application is hosted in the framework, and a phase change is warranted based on the chosen best algorithm. Real-time phase and vehicle trajectory data is fed to the application. A minimum green time duration is fixed by default, and in this paper, the minimum green time duration is 5 seconds. If the minimum duration of the green phase exceeds the current state, then phase change is warranted based on the chosen best algorithm otherwise the current state continues. The application layer communicates with the digital twin, matches the current traffic demand with the predicted traffic demand, and compares the simulation results for the parallel simulation output to choose the best algorithm. The best algorithm is chosen based on simulation performance.

| Phase | State | | | | | | | |
|---|---|---|---|---|---|---|---|---|
| | NBT | NBL | EBT | EBL | SBT | SBL | WBT | WBL |
| 0 | G | R | R | R | G | R | R | R |
| 1 | Y | R | R | R | Y | R | R | R |
| 2 | R | R | G | R | R | R | G | R |
| 3 | R | R | Y | R | R | R | Y | R |
| 4 | R | R | R | G | R | R | R | G |
| 5 | R | R | R | Y | R | R | R | Y |
| 6 | R | G | R | R | R | G | R | R |
| 7 | R | Y | R | R | R | Y | R | R |
| 8 | R | R | R | R | R | R | R | R |

**Fig. 3.** Signal phase and state.

**Algorithm DT1.** Main phase update loop for an intersection traffic signal controller based on DT1

**Input:** Subject intersection ID, **Veh_ID** - List of vehicle ID on subject intersection, **Veh_delay** - List of vehicle delay information of vehicles corresponding to vehicle ID, **Approach_ID** - List of approach connected with the subject intersection

**If** green_time <= 5
  **return**
**end if**
**for each** Group a in Approach_ID$_i$ **do**
  **for each** v in Veh_ID$_i$ **do**
    Veh_approach_delay$_i$ = Veh_delay(v) (t_out) - Veh_delay(v) (t_in)
  **end for**
  ap_delay$_i$ = ∑(Veh_approach_delay) / length(Veh_ID)
**end for**
max_avg_delay = max(ap_delay)
**if** (max_avg_delay **equals** ap_delay(NBT) **or** max_avg_delay **equals** ap_delay(SBT)) **then**
  **set** phase = 0
**else if** (max_avg_delay **equals** ap_delay(WBT) **or** max_avg_delay **equals** ap_delay(EBT)) **then**
  **set** phase = 2
**else if** (max_avg_delay **equals** ap_delay(WBL) **or** max_avg_delay **equals** ap_delay(EBL)) **then**
  **set** phase = 4
**else if** (max_avg_delay **equals** ap_delay(NBL) **or** max_avg_delay **equals** ap_delay(SBL)) **then**
  **set** phase = 6
**else**
  **signal status** = out of order
  **set** flashing yellow signal
**end if**

**Algorithm - DT2.** Main phase update loop for an intersection traffic signal controller based on DT2

**Input:** Subject intersection ID, **Veh_ID** - List of vehicle ID on subject intersection, **Veh_delay** - List of delay information of vehicles corresponding to each vehicle ID, **Approach_ID** - List of approach connected with the subject intersection, **previous_approach_veh_delay** - List of delay information of vehicles corresponding to each vehicle ID experienced in the immediate last approach

**If** green_time < =5
  **return**
**end if**
**for each** Group a in Approach_ID$_i$ **do**
  **for each** v in Veh_ID$_i$ **do**
    Veh_approach_delay$_i$ = Veh_delay(v) (t_out) - Veh_delay(v) (t_in) + previous_approach_veh_delay
  **end for**
  ap_delay$_i$ = ∑(Veh_approach_delay) / length(Veh_ID)
**end for**
max_avg_delay = max(ap_delay)
**if** (max_avg_delay **equals** ap_delay(NBT) **or** max_avg_delay **equals** ap_delay(SBT)) **then**
  **set** phase = 0
**else if** (max_avg_delay **equals** ap_delay(WBT) **or** max_avg_delay **equals** ap_delay(EBT)) **then**
  **set** phase = 2
**else if** (max_avg_delay **equals** ap_delay(WBL) **or** max_avg_delay **equals** ap_delay(EBL)) **then**
  **set** phase = 4
**else if** (max_avg_delay **equals** ap_delay(NBL) **or** max_avg_delay **equals** ap_delay(SBL)) **then**
  **set** phase = 6
**else**
  **signal status** = out of order
  **set** flashing yellow signal
**end if**

## IV. Evaluation Set-up

In this section, the Simulation of Urban Mobility (SUMO) simulation environment, baseline algorithm, vehicle delay calculation, and evaluation matrices are discussed. The TDT framework presented in this study has not been implemented and evaluated in the physical world. Instead, we simulated multiple traffic demand scenarios and evaluated the performance of the DT1 and DT2 ATSC algorithms for those scenarios. Furthermore, we tested the presented algorithms against a baseline algorithm. It is important to highlight that this study does not incorporate real-world data. The presented case study serves as a proof of concept, demonstrating the viability and potential of the proposed approach. To emulate real-world traffic flow, the SUMO is utilized. Though the simulation doesn't represent the actual TDT framework, still it can be connected to the modules of actual framework. SUMO acts as a simulation platform that encompasses the physical world, which comprises the road network, vehicles, and traffic signals. For this case study, both the physical world and DT of the physical world entities (i.e., road network DT, vehicles DT, and traffic signals DT) in the digital shadow are the same entities. Data aggregation, synchronization, and visualization are facilitated in SUMO. In the digital sibling module of the framework, the database contains real-time vehicle and traffic state data and simulation outcomes, i.e., intersection performance. To mimic various LOS, eleven traffic demands are produced and stored as SUMO demand file. The model section includes all the required models for simulating traffic flow and is already incorporated into the SUMO platform. The traffic signal is controlled using SUMO's traffic control interface (TraCI) and three algorithms: vehicle density-based baseline, delay-based DT1, and DT2, are implemented. The performance of all traffic signal control algorithms, including the baseline, DT1, and DT2, is tested against all scenarios, and the results are presented in the following section.

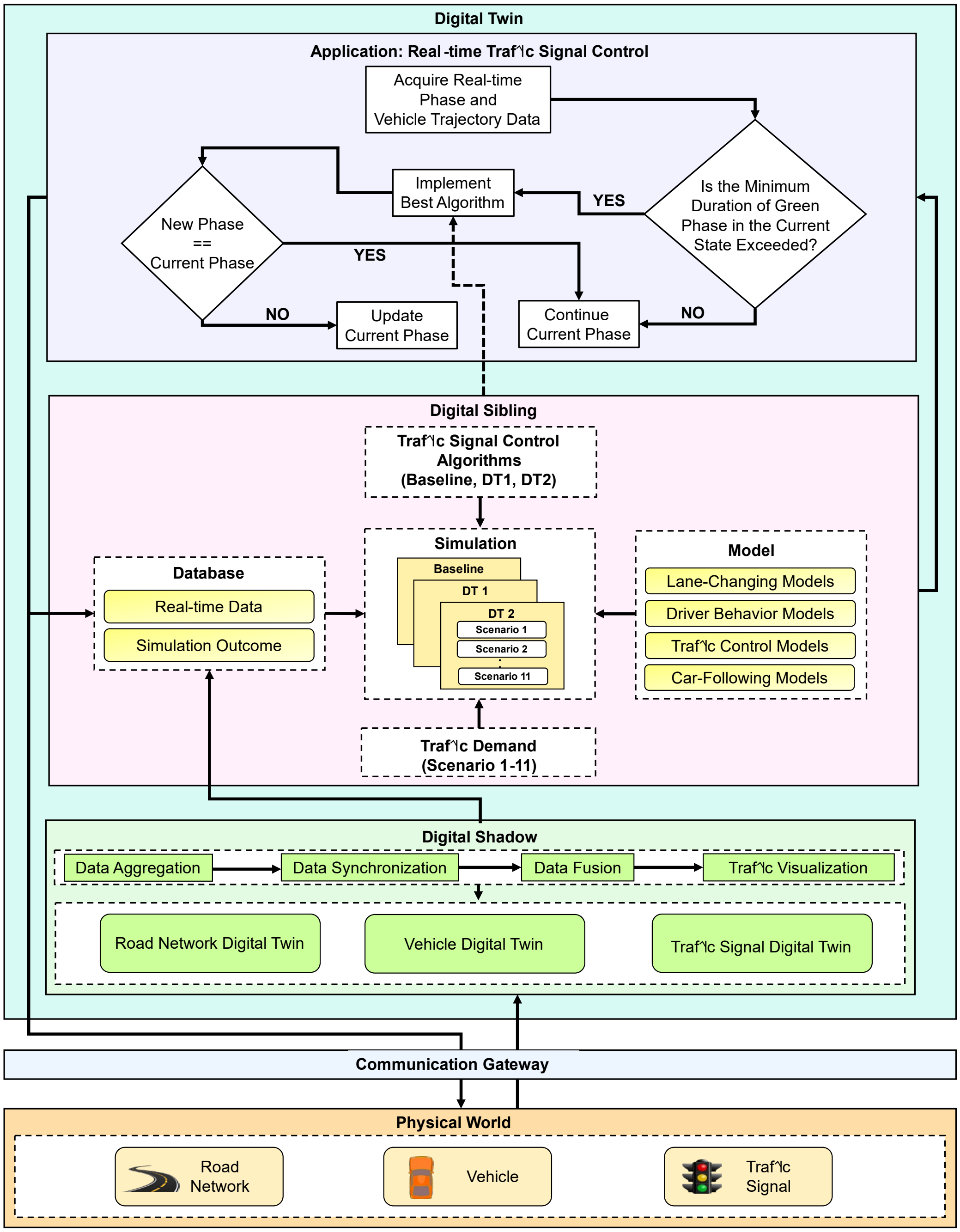

**Fig. 4.** Illustration of the delay based ATSC application framework.

### *A. Simulation Environment*

Figure 5(a) shows the SUMO graphical user interface detailing the geometry of nodes and edges. A node represents an intersection, and an edge refers to a roadway segment connecting the two nodes. Dedicated left-turn pockets are used at the junction for the left-turn movements. The connections of the possible directions a vehicle can take at the intersection are shown in Figure 5(a). The connections are the same for all the intersections in the roadway network. Figure 5(b) presents the different components of the SUMO simulation environment. The module NETEDIT is used to define the traffic demand. Multiple flows are created by selecting peripheral edges as the origin and destinations for each flow. The NETEDIT module by default uses the shortest minimum path between the origin and destination edges for flow creation and can pass through any edges in the network. Default vehicle types are used for the simulation. The simulation does not include any person or pedestrians. Different traffic demands are created by changing the vehicle generation per hour (vph) in the flow attributes. Traci is used for collecting vehicle-level data (vehicle ID, waiting time, etc.). This data is then used as inputs to the algorithm, and the traffic signal decisions are relayed back to the simulation. The traffic simulation duration is set to 3,600 seconds, including the 600 seconds of the warm-up and cool-down period each. The results reported henceforth relate to the 2,400 seconds of the actual simulation period. The traffic signals are controlled based on the baseline, DT1, and DT2 traffic signal control algorithms.

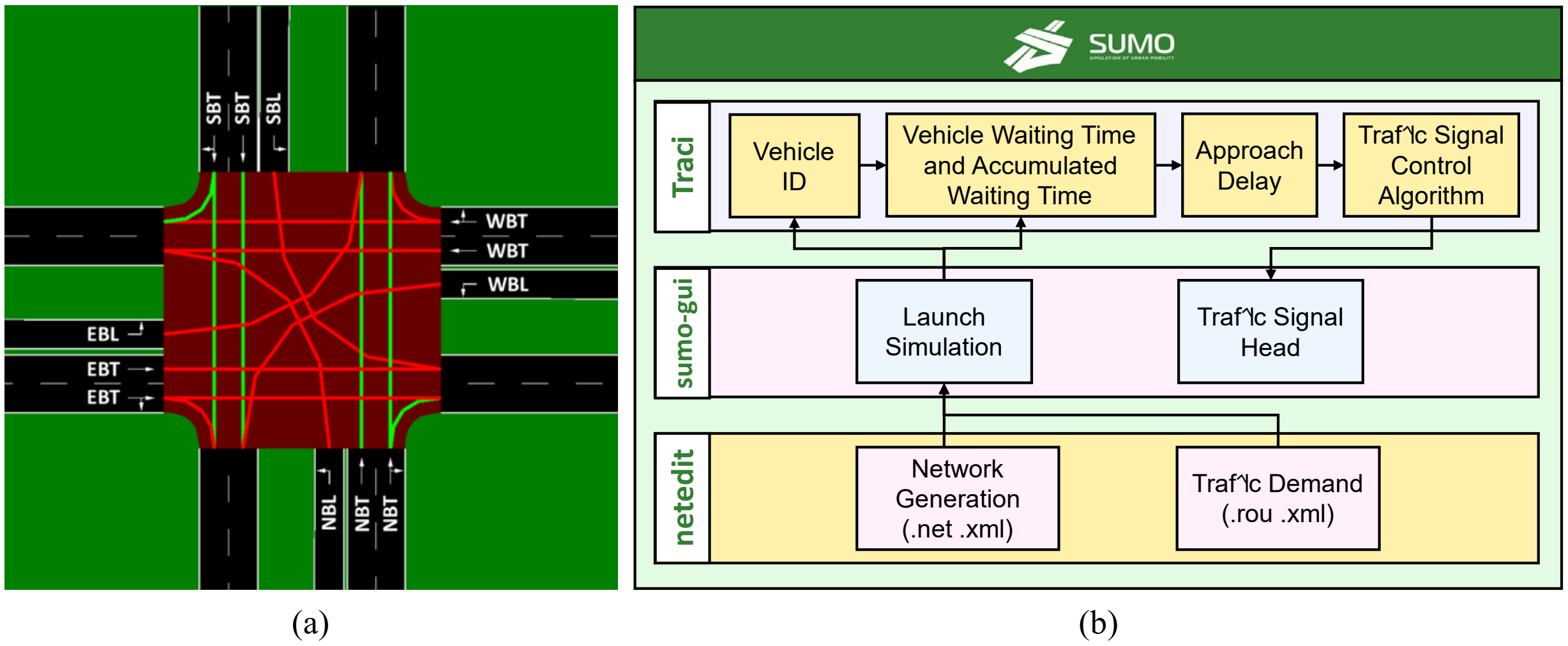

**Fig. 5.** Simulation of DT environment using SUMO (a) Connections at junctions; and (b) simulation environment.

### *B. Traffic Demand Scenarios*

For testing purpose, 10 scenarios are created varying the future traffic demands to represent different level of service. Figure 6 illustrates representative vehicle density for all eleven scenarios for a simulation setup. Notably, the vehicle density increases gradually from scenario 1 to scenario 11, which can be categorized into three groups based on their level of demand: lower demand (scenarios 1, 2, 3), moderate demand (scenarios 4, 5, 6, 7), and high demand (scenario 8, 9, 10, 11). It is important to note that the starting vehicle density remains constant for all algorithm cases, but as the simulation progresses, the density changes differently for each algorithm based on its performance.

### *C. Baseline Algorithm*

Several studies [70]–[73] have proposed density-based adaptive traffic signal control for regulating traffic at a signalized intersection. According to [73], density based ATSC minimize unnecessary waiting time and reduce the length of vehicle queues on roads. [72] concluded that density based ATSC has the potential to mitigate traffic congestion, particularly during peak hours. This, in turn, has the potential to reduce the occurrence of road accidents. [70] reported that the implementation of density-based ATSC led to a significant reduction in average delay. Because of the effectiveness of density-based ATSC, we have considered this approach as a baseline algorithm to evaluate the performance of our DT-based algorithms. The density of the approach is defined as the number of vehicles per lane per mile, calculated using Equation (5).

$$\rho_{a_i} = \frac{n_i}{n_{L_i} \times l_{L_i}} \tag{5}$$

where, $\rho_{a_i}$ is approach density of approach $i$, $n_i$ is the total number of vehicles in the subject approach, $n_{L_i}$ and $l_{L_i}$ are number of lanes in the approach $i$ and length of each lane in the approach $i$, respectively.

This section presents the pseudocode for the density-based baseline traffic signal control algorithm. The algorithm evaluates the decision for a phase change every five seconds. A phase change is initiated if the proposed or set phase is different from the current phase. When a phase change is warranted, it follows a yellow phase of two seconds and an all-red clearance of one second. Figure 3 illustrates the signal phases and their corresponding states. The algorithm does not initiate a phase change if the approach currently serving green continues to have the maximum vehicle density after 5 seconds. The green phase is assigned to the approach with the highest vehicle density, which is calculated as the number of vehicles per lane per mile. For instance, if the vehicle density on the EBT approach is highest, the algorithm will assign the green phase to the EBT approach. In case of a malfunction, the signal will flash yellow.

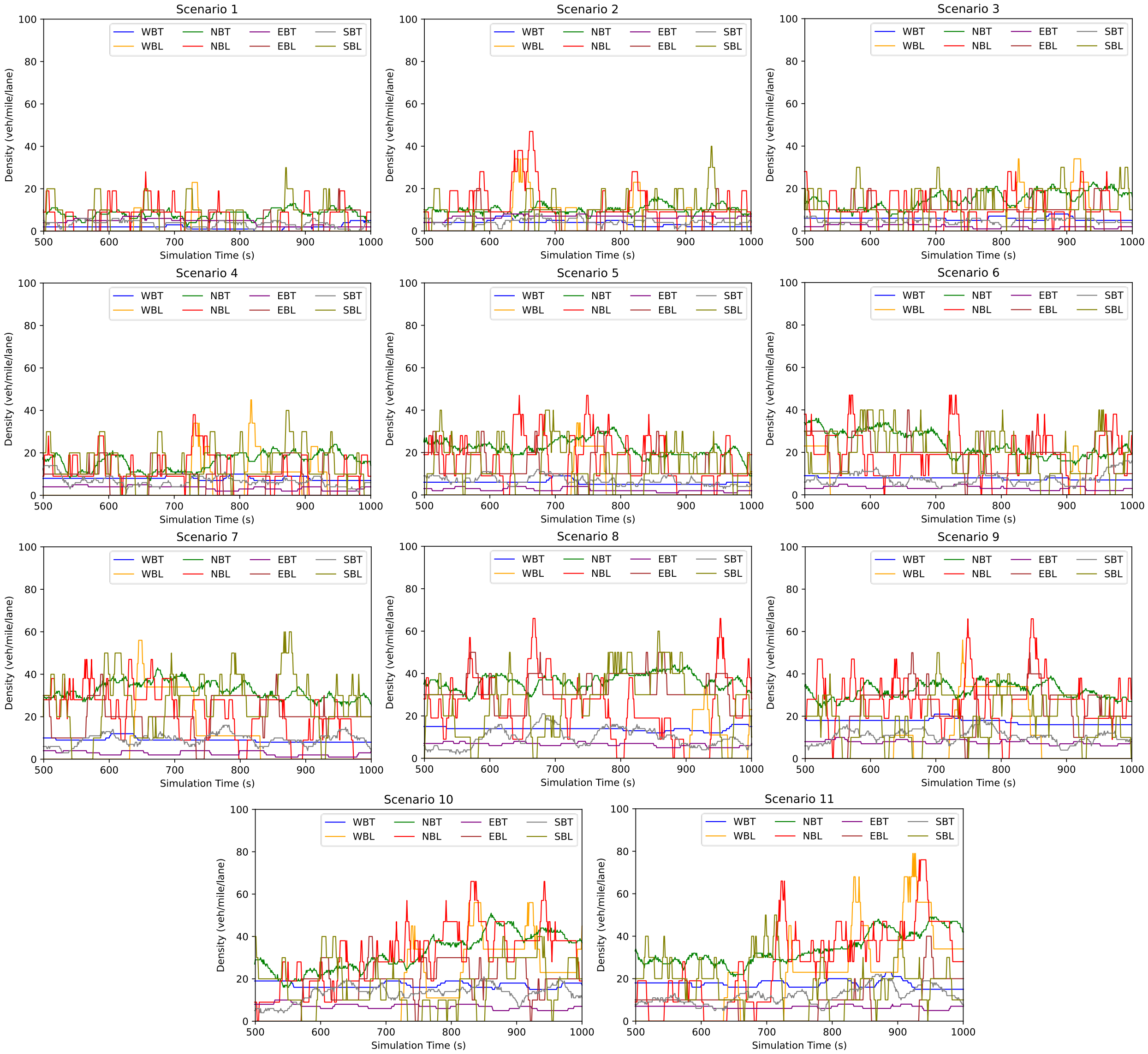

**Fig. 6.** Traffic demand scenarios (baseline).

**Algorithm baseline.** Main phase update loop for an intersection traffic signal controller based on baseline

```
Input: Subject intersection ID, Veh_ID - List of vehicle ID on subject
intersection, Veh_density - List of vehicle density of each approach
connected to subject approach
If green_time < =5
   return
end if
max_rho = max(Veh_density)
if (max_rho equals Veh_density (NBT) or max_rho equals
Veh_density (SBT)) then
     set phase = 0
else if (max_rho equals Veh_density (WBT) or max_rho equals
Veh_density (EBT)) then
     set phase = 2
else if (max_rho equals Veh_density (WBL) or max_rho equals
Veh_density (EBL)) then
     set phase = 4
else if (max_rho equals Veh_density (NBL) or max_rho equals
Veh_density (SBL)) then
     set phase = 6
else
     signal status = out of order
     set flashing yellow signal
end if
```

### *D. Vehicle Delay Calculation*

In order to measure a vehicle's delay, the Traci function getwaitingtime() is employed. According to this function, waiting time is defined as the duration in seconds that a vehicle spends at a speed below 0.1 m/s (0.22 mph) after the last time it traveled faster than 0.1 m/s [74]. Therefore, a vehicle is considered experiencing delay if its speed is 0.1 m/s or lower. However, this function has a limitation in that it initializes the delay to zero when the vehicle's speed again crosses 0.1 m/s. To obtain the cumulative delay for a vehicle, the Traci function getAccumulatedWaitingTime() is used. This function outputs the total time the vehicle's velocity is less than 0.1 m/s. Vehicle delay and accumulated delay are illustrated in Figure 7 with an example. The speed vs simulation time graph (Figure 7) shows the variation in vehicle speed at an intersection during a simulation. The primary y-axis represents the speed of the vehicles in meters per second, while the secondary y-axis displays the accumulated delay in seconds. The x-axis represents the simulation time in seconds. The graph demonstrates that there are two instances of delay, D1, and D2, during the simulation. The accumulated delay is the sum of both delays. The accumulated delay is further used to calculate the approach delay and is fed to the traffic signal control algorithms, which are explained in the subsequent sections.

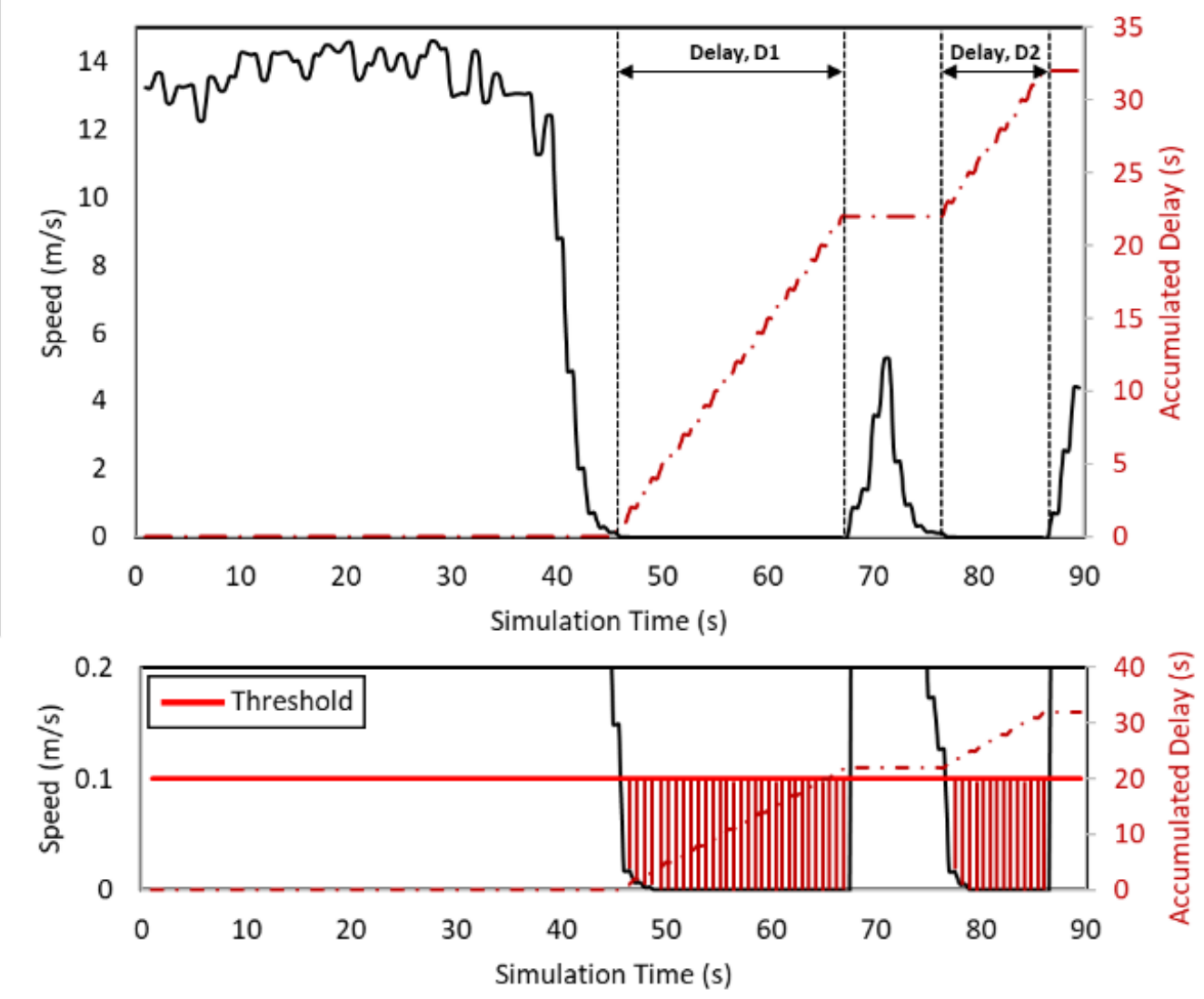

**Fig. 7.** Delay definition.

### *E. Evaluation Metrics*

This study assesses the performance of the DT1 and DT2 algorithms, as well as the baseline ATSC, at two levels (intersection and individual user) by using intersection LOS, approach average stopped delay and individual vehicle stopped delay as evaluation metrics. By evaluating the algorithms based on these metrics, we can determine their effectiveness in improving the traffic flow and reducing congestion at the intersection.

**Level of service (LOS):** The Highway Capacity Manual [75] describes LOS as a subjective measure that characterizes the operational conditions within a traffic stream and how they are perceived by drivers or passengers. For signalized intersections, LOS is defined based on the average control delay and ais shown in Table 2. The letter grades A to F are used to designate the LOS of an intersection, with A indicating the best operating conditions and F indicating the worst. Details of LOS calculation provided in the next section. The details of LOS calculation are presented in the next section.

**Approach Average Stopped Delay (AASD):** This metric is to calculate the average stopped delay experienced by all the vehicles in through and left turn approaches. AASD helps in assessing the impact of the ATSC algorithms for reducing delays at the approach level and represents intersection performance.

**Distribution of Stopped Delay (DSD):** DSD metric is defined as the delay experienced by individual vehicle at approached connected to the subject intersection. The delay distribution for each movement can be analyzed in terms of frequency to compare the effectiveness of different signal control strategies. The skewness of the delay frequency distributions can be compared to determine the superiority of one strategy over the others. This metric provides insight into how the delay is distributed among different road users and helps in identifying the road user experience which passing through the subject intersection.

TABLE 2
HCM LEVEL OF SERVICE CRITERIA FOR SIGNALIZED INTERSECTIONS [75]

| **LOS** | **Average Control Delay (s/veh)** |
|---|---|
| A | 10 |
| B | > 10 and 20 |
| C | > 20 and 35 |
| D | > 35 and 55 |
| E | > 55 and 80 |
| F | > 80 |

## V. Results and Discussion

This study assesses the performance of the DT1 and DT2 algorithms, as well as the baseline ATSC, at two levels (intersection and individual user) by using the above-mentioned evaluation metrics. All three algorithms are ATSC algorithms.

### *A. Intersection Performance: Intersection Level*

Figure 8 shows the subject intersection LOS for baseline, DT1, and DT2 algorithms. The LOS for each approach and the intersection is determined by using the control delay. Control delay measures the total delay that a vehicle experiences when it moves through a part of the road that is impacted by a traffic signal [76]. The delay includes the time taken for deceleration, the time spent while stopped, and the delay experienced while accelerating to reach the running speed. Since control delay is a more reliable indicator of user delay compared to other delays, it has been used by the Highway Capacity Manual (HCM) for calculating the LOS. SUMO default waiting time function is defined as the duration (in seconds) that a vehicle spends at a speed below 0.1 m/s (0.22 mph) after the last time it traveled faster than 0.1 m/s (“Python: module traci._vehicle,” n.d.). Hence, the SUMO default waiting time function cannot be used to calculate the control delay as the waiting time function can only provide the stopped delay and there is no way to calculate the acceleration and deceleration delay. Therefore, the segment delay is used as an alternative for calculating control delay, which has been found to be a reliable predictor according to [76]. The segment delay is measured in seconds per vehicle and is defined as the difference between the total travel time of a vehicle in a segment and the free-flow travel time for that segment. This includes any delay experienced by a vehicle due to speed reduction or complete stops caused by traffic signals or other factors within the segment. Therefore, in order to determine the segment delay for each vehicle for an approach between two consecutive intersections, the time spent by each vehicle in that approach is subtracted from the free flow travel time. The following equation is used to calculate the segment delay, $D_{sij}$; for ith vehicle and jth approach;

$$D_{sij} = (t_o - t_i) - \frac{L}{v_{ff}} \tag{6}$$

Where, $t_o$, time the ith vehicle exists the jth approach; $t_i$, time the ith vehicle enters the jth approach; $L$, length of the jth approach; $v_{ff}$, free flow speed.

algorithms. As the primary objective of DT-based ATSC approaches is to reduce the control delay of the subject intersection, based on the presented results, it is evident that DT1 and DT2 algorithms can significantly reduce the control delay at an intersection and further enhance user experience. The details of the individual user experience improvement are presented in subsection *V(C)*.

### *B. Approach Performance : Intersection Level*

Figure 9 depicts the average delay in the subject intersection for various through movements in scenario 3, which was chosen randomly. Here average delay is the average stopped delay experienced by the vehicles while

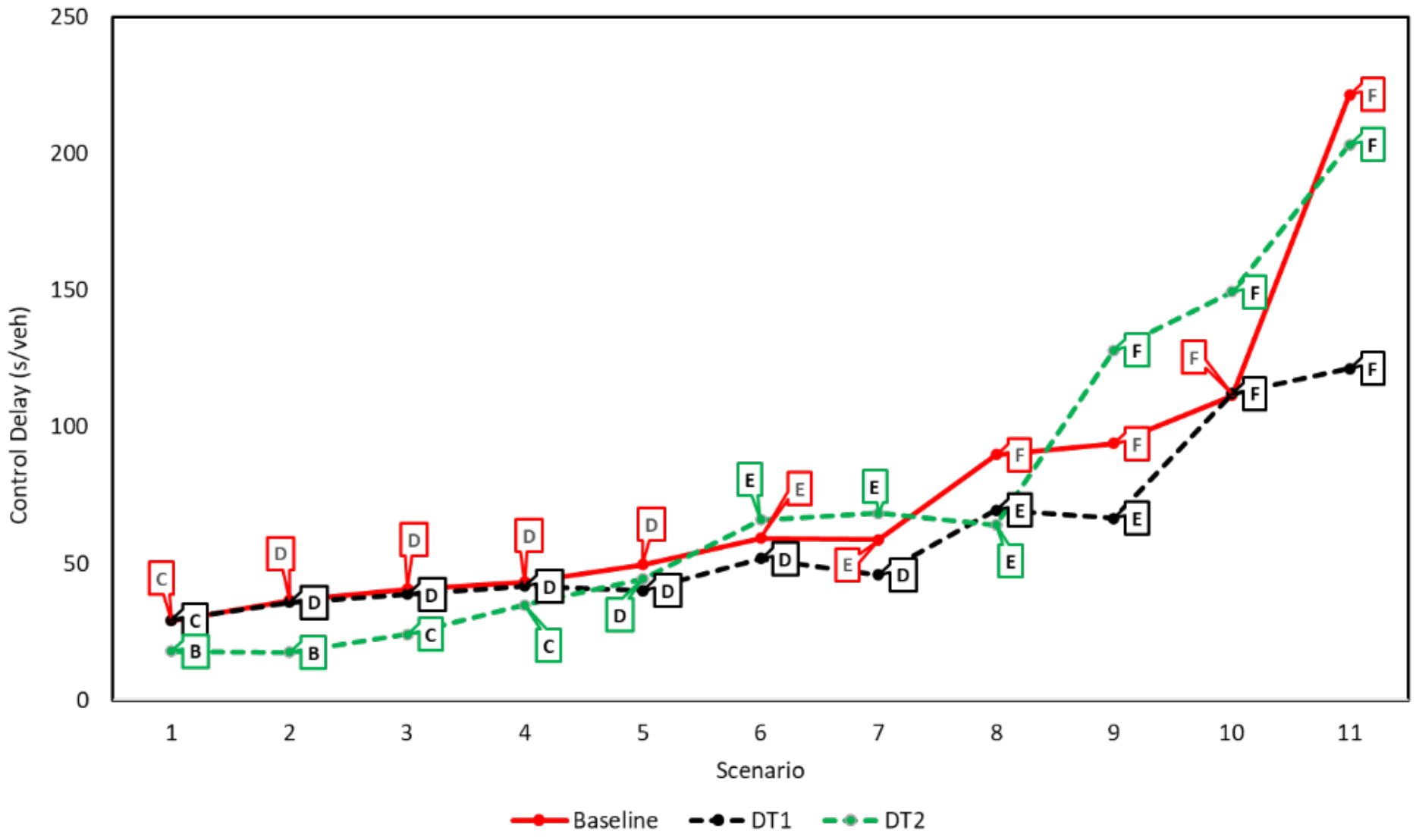

**Fig. 8.** Subject intersection LOS

Figure 8 demonstrates the performance of DT1, DT2, and the baseline algorithm across different demand scenarios. For lower demands (scenarios 1, 2, and 3), DT2 outperforms both the baseline and DT1 algorithms and for moderate and higher demands DT1 performs better than DT2. While DT1 and the baseline algorithm achieve the same LOS in scenarios 1, 2, 3, 4, and 5, DT1 consistently exhibits a 1% to 19% reduction in control delay compared to the baseline algorithm. Notably, in scenarios 6, 7, and 8, DT1 surpasses the baseline algorithm with a control delay reduction ranging from 13% to 29%. These results indicate that DT1 performs better as demand increases. In scenario 10, both DT1 and the baseline algorithm perform equally and fail to meet the desired control delay. However, in scenario 11, while both algorithms fail, DT1 significantly reduces the control delay compared to the baseline algorithm. For scenarios 1, 2, 3, and 4, DT2 consistently outperforms the baseline algorithm by reducing the control delay by 19% to 52%. On the other hand, in scenarios 5 and 6, although there is no improvement in terms of LOS, DT2 successfully reduces the control delay by 10%. Note that in scenarios 6, 7, 9, 10, and 11, DT2 performs worse than the baseline algorithm, leading to failure for both passing through the approach connected to the subject intersection. The results indicate that the baseline ATSC outperforms the DT1 and DT2 ATSCs for the NBT and SBT movements; however, performs poorly for the EBT and WBT movements. Specifically, the mean of the average delay for NBT, SBT, EBT, and WBT movements for the entire simulation are 1.57, 0.73, 46.47, and 59.54 seconds, respectively, for the baseline. In contrast, for DT1, the mean of the average delay for the same movements are 6.53, 4.8, 6.12, and 6.25 seconds, respectively, and for DT2, the mean of the average delay for the same movements are 2.87, 1.88, 2.55, and 3.82 seconds, respectively. Although the baseline outperforms DT1 and DT2 for the NBT and SBT movements, it results in a significantly higher average delay for the EBT and WBT movements. In contrast, the DT1 and DT2 algorithms distribute demand in a balanced manner and serve all the through movements in an optimized fashion, leading to a lower overall average delay. Furthermore, the DT2 ATSC significantly reduces the delay at the subject intersection for scenario 3. These results highlight that incorporating information on the subject and upstream movements can reduce the average delay for an intersection.

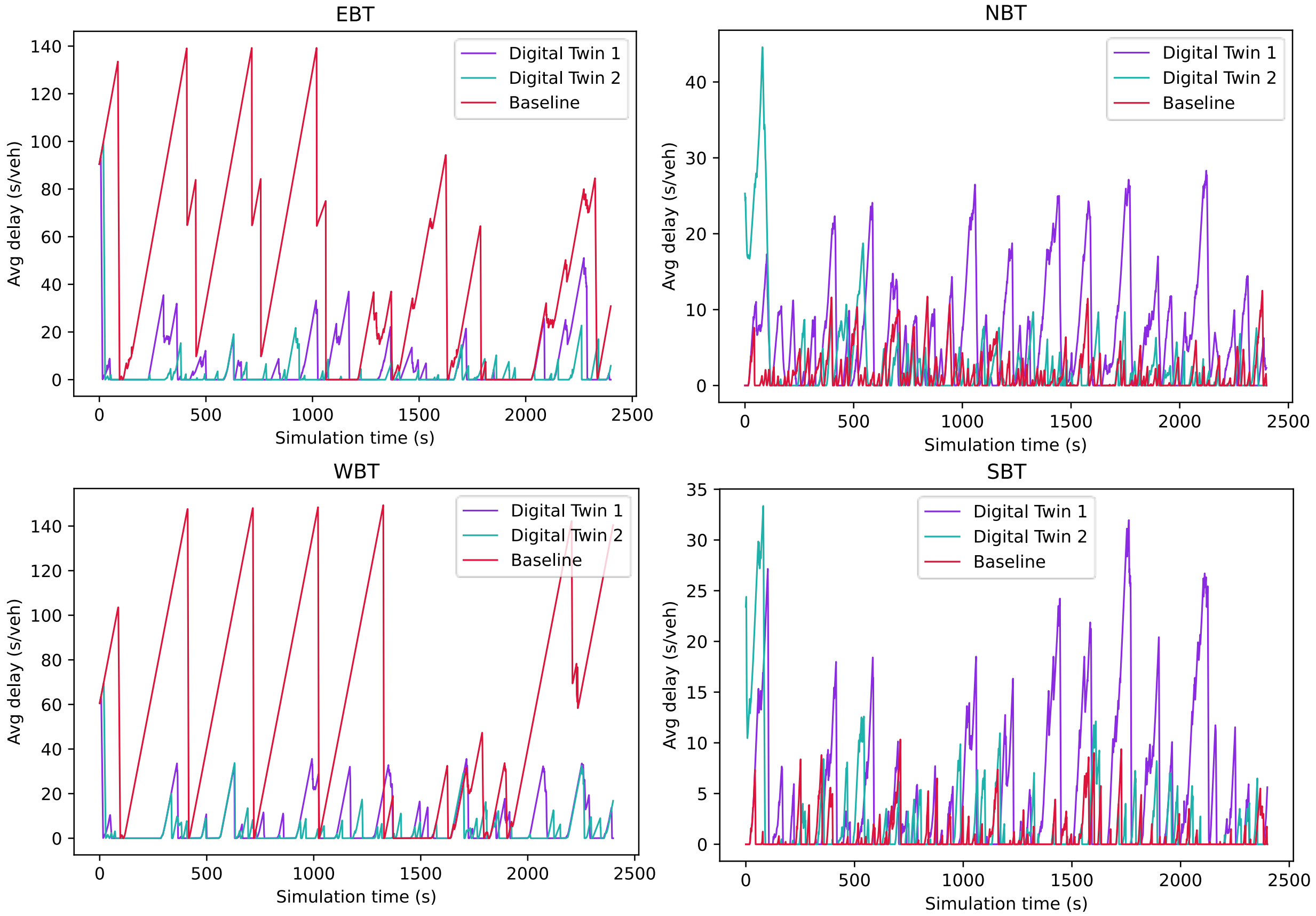

**Fig. 9.** Average delay profiles for the dedicated through movements at the subject intersection (Scenario 3).

Figure 10 displays the average delay values for the left-turn movements at the subject intersection in scenario 3. The baseline ATSC algorithm has a mean delay of 2.94, 2.02, 10.78, and 3.71 seconds for the NBL, SBL, EBL, and WBL movements, respectively. In contrast, the DT1 algorithm has a mean delay of 13.21, 12.48, 11.68, and 2 seconds for the same movements, while the DT2 algorithm has a mean delay of 4.68, 8.26, 4.61, and 0.148 seconds respectively. The DT2 algorithm performs better for the EBL and WBL movements, whereas the baseline algorithm performs best for the NBL and SBL movements due to the higher vehicle density in those movements. In conclusion, the DT2 algorithm outperforms the other algorithms for left-turn movements at the subject intersection in scenario 3.

*C. User Experience : Individual User Level*

The distribution of stopped delays of vehicles at the subject intersection for all approaches in scenario 3 is illustrated in Figure 11. The x-axis represents stopped delay at subject intersection in seconds, while the y-axis shows the frequency of the vehicles in each bin. Results for baseline, DT1 and DT2 algorithms are represented by salmon pink, purple and green bars respectively. As we can see, for all the approaches the distributions are highly right skewed. It can be observed that the distributions for all approaches are strongly skewed to the right. This indicates that the three ATSC algorithms have yielded more vehicles with lower stopped delay. The EBT and WBT approaches exhibit a notable difference in the distribution of vehicle delays between DT1 and DT2 compared to the baseline case. Specifically, for DT1 and DT2, the majority of vehicles experience lower levels of delay, as indicated by the high frequency of the lowest delay bin. In contrast, the baseline case has fewer vehicles with low delay and a noticeable number of vehicles experiencing very high delays. On the other hand, the performance of the approaches NBT and SBT shows that the baseline algorithm performs the best, with all vehicles experiencing the smallest delay. In contrast, for DT1 and DT2, delays are more evenly distributed across all bins with higher delays. However, a closer examination of the delay values reveals that the highest delays fall within the lowest delay bin for both EBT and WBT approaches. Therefore, in terms of through movements, DT2 outperforms DT1 by more effectively distributing the delays compared to the baseline algorithm, and both ATSC algorithms outperform the baseline.

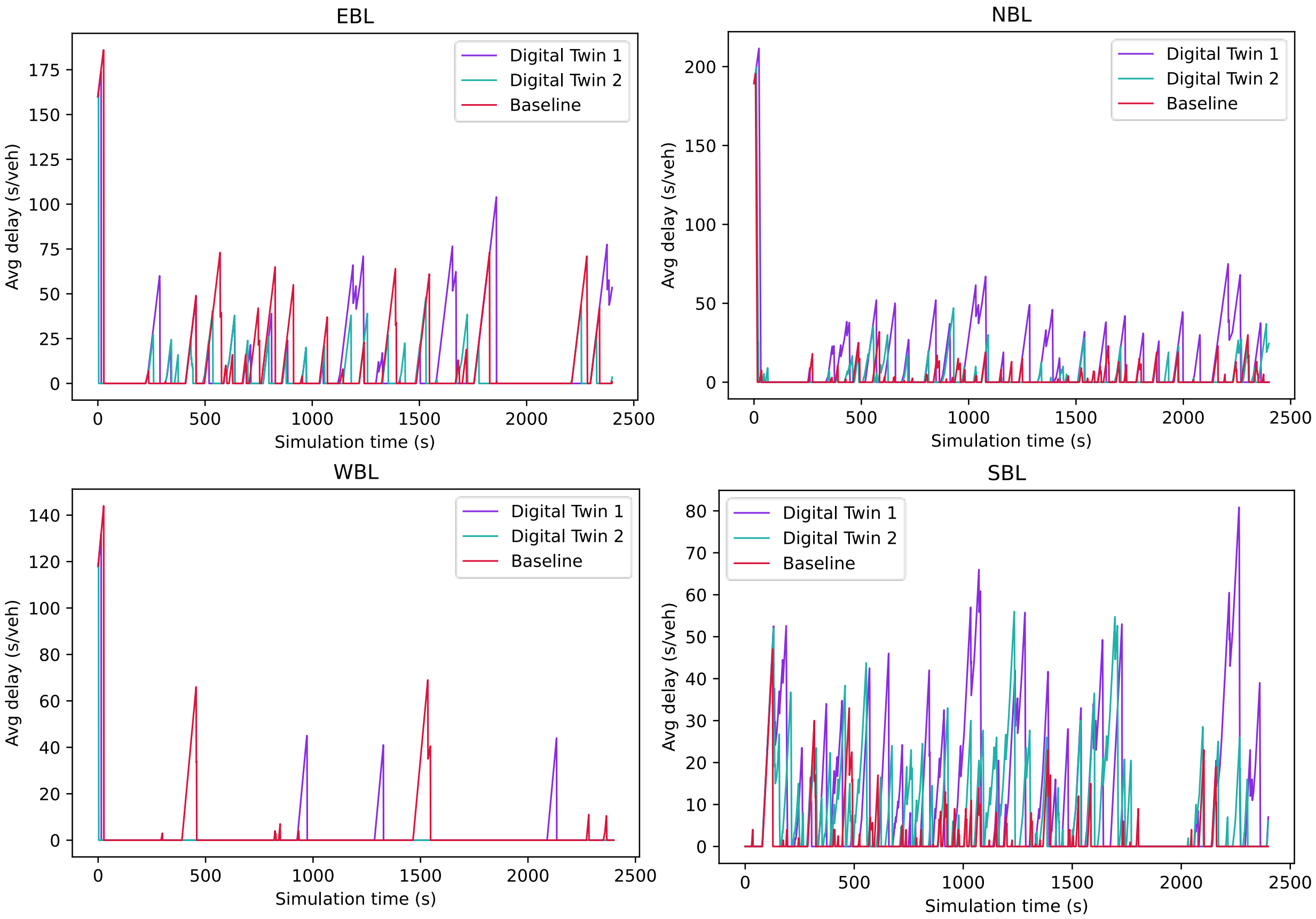

**Fig. 10.** Average delay profiles for the dedicated left turn movements at the subject intersection (Scenario 3).

In the case of left turn movements, in EBL approach, DT2 outperforms both DT1 and the baseline algorithm with a highly right skewed delay distribution. On the other hand, for DT1, delays are more evenly distributed, but there are more vehicles with high delays. The baseline algorithm performs in between DT1 and DT2 for left turn movements. The baseline algorithm performs the best in WBL, NBL and SBL approaches. DT1 outperforms DT2 in the SBL approach, while DT2 performs better than DT1 in the WBL approach. The NBL approach shows that DT1 results in a more evenly distributed delay compared to DT2.

## VI. Future Research Direction

This paper has presented a DT-based ATSC framework and evaluated its performance through the implementation of two ATSC algorithms at a subject intersection in a network. The next step is to implement the algorithms networkwide and assess their impact on road user experience throughout the network. Figure 12 illustrates an example of such an implementation, with solid light blue lines representing all the approaches within the network and representing approaches whose delay data is utilized by the ATSC algorithm to make traffic signal control decisions. The solid black line indicates the path of a particular vehicle, while the solid red and empty green circles represent the subject and contributing intersections, respectively. The subject intersection is the main intersection under consideration, while the contributing intersection is the previous intersection whose data is used in the DT2 algorithm. Figures 12(A) and 12(B) illustrate the execution of DT1 and DT2 algorithms in a single intersection of the network. Our future research objective is to extend the implementation of DT1 and DT2 to all intersections in the network, where every intersection is the subject intersection denoted by the solid red circles in Figure 12(C). This network-wide deployment of the DT-based ATSC framework is expected to balance the delay experienced by all vehicles travelling through the network. From the perspective of a single vehicle, Figures 12(D) and 12(E) demonstrate the implementation of DT1 and DT2 algorithms, respectively, as the vehicle moves from one intersection to another. The delay-based algorithms should reduce delay in later intersections if higher waiting times are experienced in earlier intersections, or balance delays across the network to reduce overall delays. The use of DT-based ATSC algorithms can lead to less delay for all vehicles in the network. Moreover, new ATSC algorithms can be tested using the proposed framework.

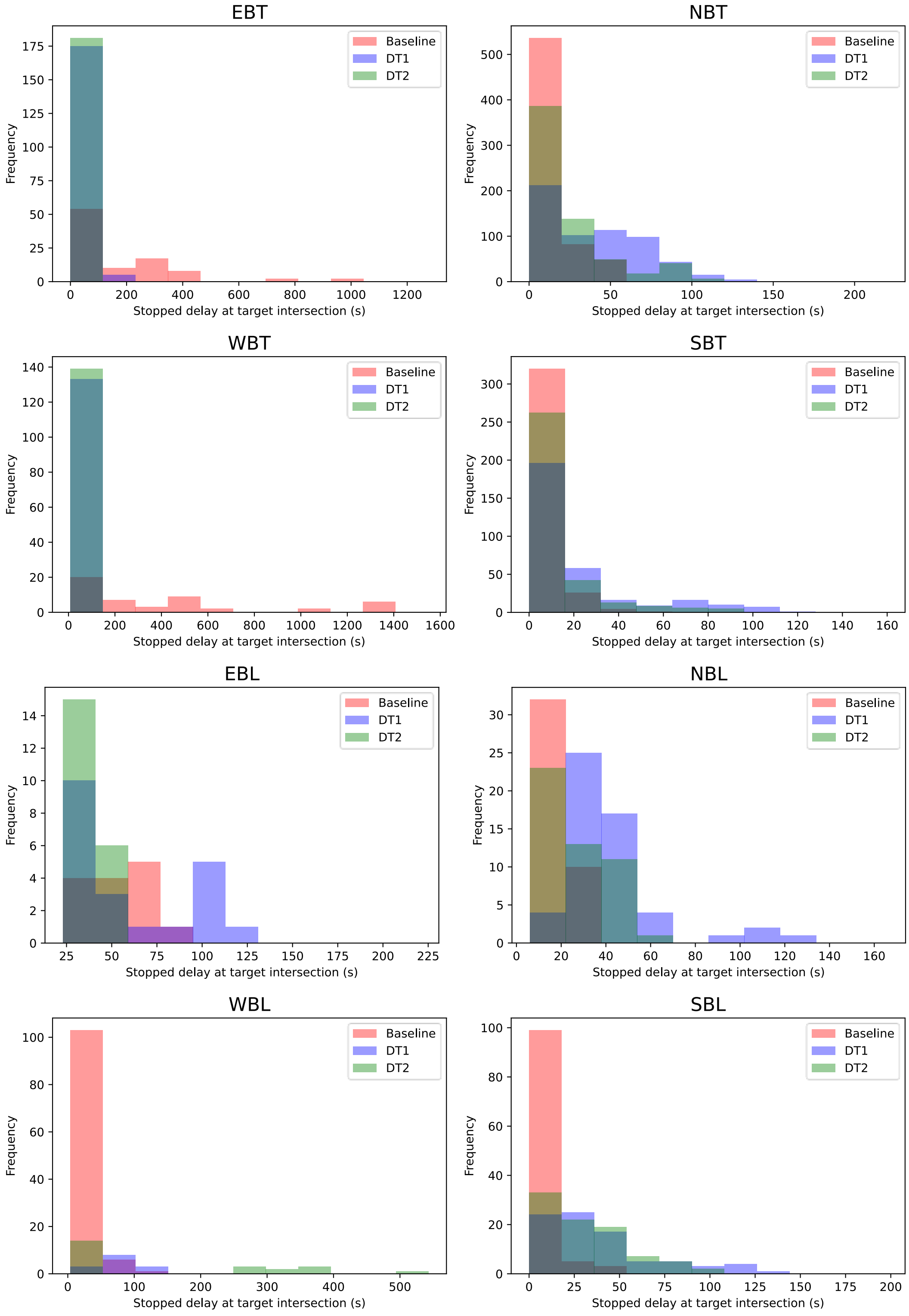

**Fig. 11.** Stopped delay distribution for all vehicles at the subject approaches (Scenario 3).

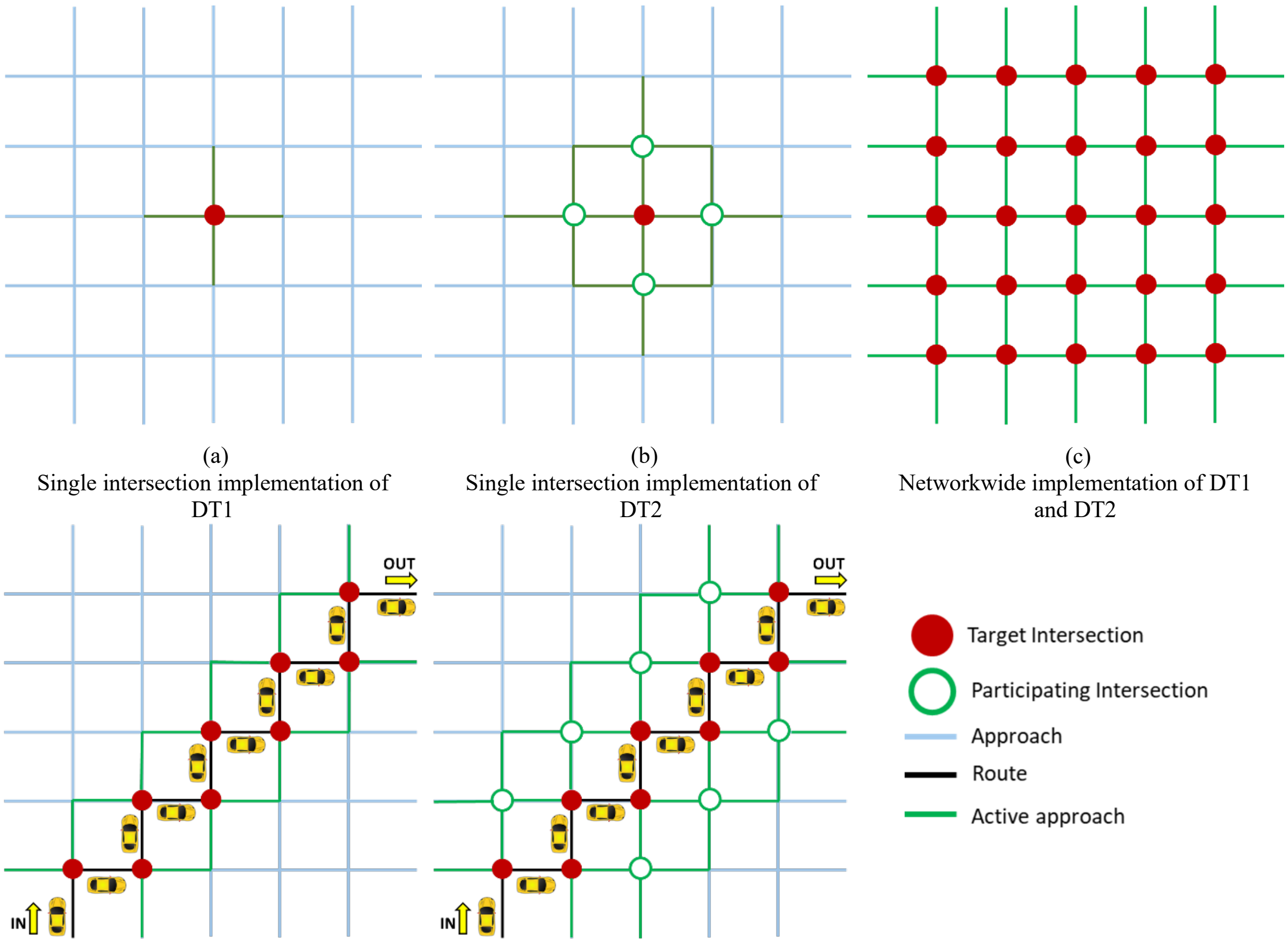

(a) Single intersection implementation of DT1

(b) Single intersection implementation of DT2

(c) Networkwide implementation of DT1 and DT2

(d) Networkwide implementation of DT1 single vehicle perspective

(e) Networkwide implementation of DT2 single vehicle perspective

**Fig. 12.** Networkwide implementation.

## VII. Conclusion

In this study, a comprehensive ATSC framework was introduced, leveraging TDT technology to enhance signalized intersection performance and user satisfaction. By utilizing real-time individual vehicle trajectory data, considering future traffic demands and employing a parallel simulation strategy, two DT-based ATSC algorithms, DT1 and DT2, were developed. DT1 considered the delay of each vehicle from all approaches associated with the subject intersection, while DT2 extended its analysis to include the delay of vehicles from both the subject intersection and the immediate adjacent intersection. To evaluate the effectiveness of these algorithms, various traffic demand scenarios were examined, ranging from low to high levels, at both the intersection and individual user levels.

The evaluation results clearly demonstrated the superiority of DT1 and DT2 over the density-based baseline approach in terms of reducing control delays. For low traffic demands, both DT1 and DT2 achieved significant control delay reductions ranging from 1% to 52%. Additionally, DT1 outperformed DT2 in moderate traffic scenarios, achieving reductions of 3% to 19%, while DT2 still achieved notable reductions of 10% to 19% compared to the baseline approach under high traffic demand. Notably, DT1 achieved control delay reductions of 1% to 45% in high traffic scenarios, whereas DT2 achieved reductions of 8% to 36% compared to the baseline algorithm. Moreover, both DT1 and DT2 effectively distributed delays among all vehicles, in contrast to the baseline ATSC algorithm. This improved distribution of delays played a key role in enhancing user satisfaction. The findings highlight the significant advantages of employing DT1 and DT2 in reducing control delays and improving overall signalized intersection performance. Consequently, the choice between DT1 and DT2 can be determined based on specific traffic demands, providing a tailored approach to traffic signal control optimization. The future goal is to implement the algorithms networkwide and assess their impact on road user experience. The framework aims to balance delays across the network and reduce overall delays for all vehicles. Additionally, the proposed framework allows for testing of new ATSC algorithms.

## Acknowledgment

This material is based on a study supported by the Alabama Transportation Institute (ATI). Any opinions, findings, conclusions, or recommendations expressed in this material are those of the author(s) and do not necessarily reflect the views of the Alabama Transportation Institute (ATI).